# Complete reactants-to-products observation of a gas-phase chemical reaction with broad, fast mid-infrared frequency combs

Nazanin Hoghooghi[1,*,†,], Peter Chang[2,3,†], Scott Egbert[1], Matt Burch[4], Rizwan Shaik[4], Patrick Lynch[4], Scott Diddams[2,3,5], and Gregory B. Rieker[1,*]

[1]Precision Laser Diagnostics Laboratory, University of Colorado, Boulder, CO
[2]Time and Frequency Division, National Institute of Standards and Technology, Boulder, CO
[3]Department of Physics, University of Colorado, Boulder, CO
[4]Department of Mechanical and Industrial Engineering, University of Illinois Chicago, IL
[5]Electrical, Computer and Energy Engineering Department, University of Colorado, Boulder, CO

*Corresponding authors: nazanin.hoghooghi@colorado.edu, greg.rieker@colorado.edu
†These authors contributed equally.

**Abstract**
Molecular diagnostics are a primary tool of modern chemistry, enabling researchers to map chemical reaction pathways and rates to better design and control chemical systems. Many chemical reactions are complex and fast, and existing diagnostic approaches provide incomplete information. For example, mass spectrometry is optimized to gather snapshots of the presence of many chemical species, while conventional laser spectroscopy can quantify a single chemical species through time. Here we optimize for multiple objectives by introducing a high-speed and broadband, mid-infrared dual frequency comb absorption spectrometer. The optical bandwidth of >1000 cm$^{-1}$ covers absorption fingerprints of many species with spectral resolution <0.03 cm$^{-1}$ to accurately discern their absolute quantities. Key to this advance are 1 GHz pulse repetition rate frequency combs covering the 3-5 µm region that enable microsecond tracking of fast chemical process dynamics. We demonstrate this system to quantify the abundances and temperatures of each species in the complete reactants-to-products breakdown of 1,3,5-trioxane, which exhibits a formaldehyde decomposition pathway that is critical to modern low temperature combustion systems. By maximizing the number of observed species and improving the accuracy of temperature and concentration measurements, this spectrometer advances understanding of chemical reaction pathways and rates and opens the door for novel developments such as combining high-speed chemistry with machine learning.

**Introduction**
Better understanding of chemical reaction pathways and rates has enabled large gains in transportation fuel efficiency, lower emissions chemical manufacturing, improved fire suppression, and hypersonic propulsion, among other innovations. Oversimplification or oversight with respect to even minor pathways can have negative implications on models and their macroscopic predictions of reaction behavior, as recently observed with roaming reactions, termolecular reactions, and non-thermal prompt reactions (*1–4*). Controlled chemical reactors together with *in situ* diagnostics are key to improving understanding of these pathways. An ideal *in situ* diagnostic approach would reveal the complex reaction pathways and kinetic rates by 1) probing *all* critical chemical species from reactants to products, 2) with high specificity and concentration accuracy, 3) at the time scales of the reaction (often very short).

Here we introduce and demonstrate a dual frequency comb absorption spectrometer that achieves all three desirable characteristics for chemical kinetic diagnostics by providing broad optical bandwidth to measure the mid-infrared absorption fingerprint of complex molecular reactants with simultaneously high resolution to accurately distinguish small intermediate and product species and rotational/vibrational temperatures, at microsecond timescales.

The invention of broad optical bandwidth mode-locked frequency combs opened new potential for accurate, multi-species laser-based diagnostics. Quantum energy level transitions of many molecules could be probed simultaneously with thousands of finely spaced and precisely known optical frequencies that can be individually measured at comb tooth resolution by interfering two combs together, a technique called dual-comb spectroscopy (DCS) (*5, 6*). DCS broke the size/acquisition-rate/resolution limitations of the Fourier transform spectrometers (FTS) that have been a workhorse of optical absorption sensing for decades, replacing the mechanical scanning arm of the FTS Michelsen interferometer with the motionless interference of the two frequency combs. However, it has still been challenging to perform DCS over broad optical bandwidth with high resolution at the microsecond acquisition speeds necessary to study many physical processes because of inherent limitations among frequency comb types.

The maximum optical bandwidth ($\Delta \nu$) achievable with DCS for a given acquisition time ($\tau_{ac}$) is given by $\Delta \nu / \tau_{ac} \leq f_{rep}^2 / 2$, where $f_{rep}$ is the spacing between comb modes (named for the pulse repetition rate of mode-locked lasers, which matches the mode spacing) (*6*). Larger frequency comb mode spacing greatly improves bandwidth. However, the spectral point spacing (resolution) of the resultant spectrum decreases with higher mode spacing. In addition, not all frequency comb architectures are able to generate light across the full detectable optical bandwidth given by the equation above. This issue is visualized in Figure 1 for the three primary *mid-infrared* frequency comb architectures. Quantum cascade and interband cascade frequency combs (QCL/ICL) have 10+ GHz comb mode spacing that achieves high acquisition speed and high power-per-mode, though with limited single-shot resolution and optical bandwidth (*7–11*). Electro-optic modulator combs can have narrow mode spacing (e.g. 300 MHz) to achieve high resolution, but with limited optical bandwidth(*12–14*). Finally, mode-locked combs that generate large mid-infrared optical bandwidth do so with lower pulse repetition rate to maintain high pulse power to drive the nonlinear optical processes that create the bandwidth (*15–21*).

The spectrometer described here is enabled by mode-locked frequency combs that achieve both broad bandwidth and high power-per-comb-mode in the mid-infrared at 1 GHz pulse repetition rates using chirped pulse amplification and a robust intra-pulse difference frequency generation technique (*20, 22*). The spectrometer spans the entire C-H stretch rovibrational energy range with no gap (3-5 μm, >1000 cm$^{-1}$), spectral point spacing of 0.03 cm$^{-1}$ (1 GHz), and acquisition rate of up to 167 cm$^{-1}$/10 μs per detector (the maximum DCS bandwidth for $\tau_{ac}$=10 μs and $f_{rep}$ =1 GHz as described in the DCS equation above). The 1 GHz mode spacing and broad bandwidth can resolve most molecular species at the temperatures, pressures, and timescales of reacting systems, particularly as multiple detectors are employed.

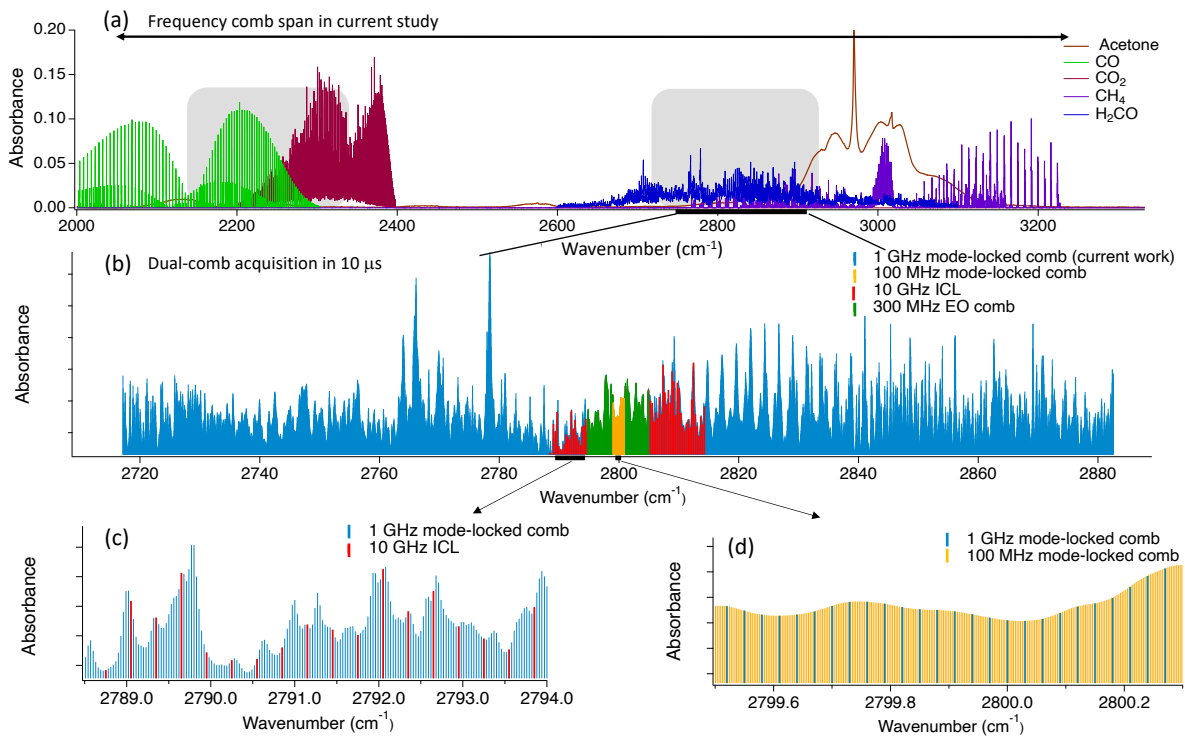

**Figure 1. Comparing mid-infrared frequency comb spectrometers.** (a) Continuous span of the 1 GHz repetition rate dual-comb spectrometer introduced in this study with example underlying molecular absorption spectra. The two shaded regions of the spectrum are simultaneously acquired in 17.5 µs by the two-detector configuration of this study. (b) Span that can be acquired in 10 µs in dual-comb configuration for various mid-infrared frequency comb architectures. (c-d) Zoom of single-shot spectral resolution possible with various mid-infrared dual-comb architectures.

With this new capability, we study the decomposition of 1,3,5-Trioxane ($C_3H_6O_3$) to formaldehyde ($H_2CO$), and formaldehyde's subsequent reactions to carbon monoxide (CO) in a high-repetition rate shock tube (*23*). Formaldehyde is known to be a critical intermediate species in low temperature combustion en route to ignition (*24–27*). The strong endothermicity of the precursor decomposition further complicates this system, which leads to uncertainty in rate constants from the uncertain temperature assignment following the decomposition of the precursor (*10, 28*). We simultaneously measure time-resolved concentrations of molecular reactants, smaller intermediates, products (CO), and the gas temperature throughout the reaction with high accuracy and 17.5 µs time resolution. Quantitatively tracking an array of molecular species and including temperature constrains reaction pathways and rates. This molecular 'cinema' can provide new insight into a host of standing kinetic research challenges.

**Experiments**

Among the mid-infrared frequency comb architectures, mode-locked combs have demonstrated the broadest optical output bandwidth due to the ability to leverage high peak pulse power to drive nonlinear frequency generation and conversion processes. However, mid-infrared mode-locked combs have thus far been demonstrated at pulse repetition rates ≤200 MHz, which limits the DCS bandwidth to 6.7 cm$^{-1}$/10µs – enough to probe only a single absorption transition of a small molecule at microsecond timescales. Here, we improve the optical bandwidth by two orders of magnitude, enough to measure overlapping spectra of multiple large and small molecules at microsecond timescales, by creating a mid-infrared dual frequency comb spectrometer with 1 GHz repetition rate. Increasing repetition rate reduces peak pulse power (for a given average laser power), which hinders the nonlinear processes. The breakthrough that led to the spectrometer was the generation of sub-two cycle (8 fs) pulses at 1 GHz repetition rate, centered at 1.56um with ~0.25 MW peak power (*22*), which efficiently drives the IP-DFG processes to generate the broadband mid-infrared frequency comb. We generate these pulses using chirped pulse amplification (CPA) and soliton self-compression in an anomalous dispersion highly nonlinear fiber(*29*). These pulses then drive the IP-DFG inside a $\chi^{(2)}$ nonlinear crystal to generate mid-infrared light. We use a fan-out periodically poled lithium niobate (PPLN) crystal. The result is a broadband 1 GHz MIR frequency comb, with > 4 mW of power and continuous coverage of the entire 3-5 μm region. Both GHz frequency combs are fully stabilized to enable simple coherent averaging of the dual-comb signal (see Supplemental).

The acquisition rate of the spectrometer is defined by the difference between the repetition rates of the two frequency combs and can be adjusted from a few Hz up to 100 kHz (the limit of the piezo actuator range that adjusts the cavity length of the near-infrared oscillator). After passing through the experiment, we split and filter the frequency comb light into two wavelength regions, each passed to a separate detector (Figure 2 (a)). This doubles the optical bandwidth for a given acquisition rate (*30*), and improves the SNR by reducing the bandwidth of light on each detector to provide more power-per-comb-mode before detector saturation (*31*).

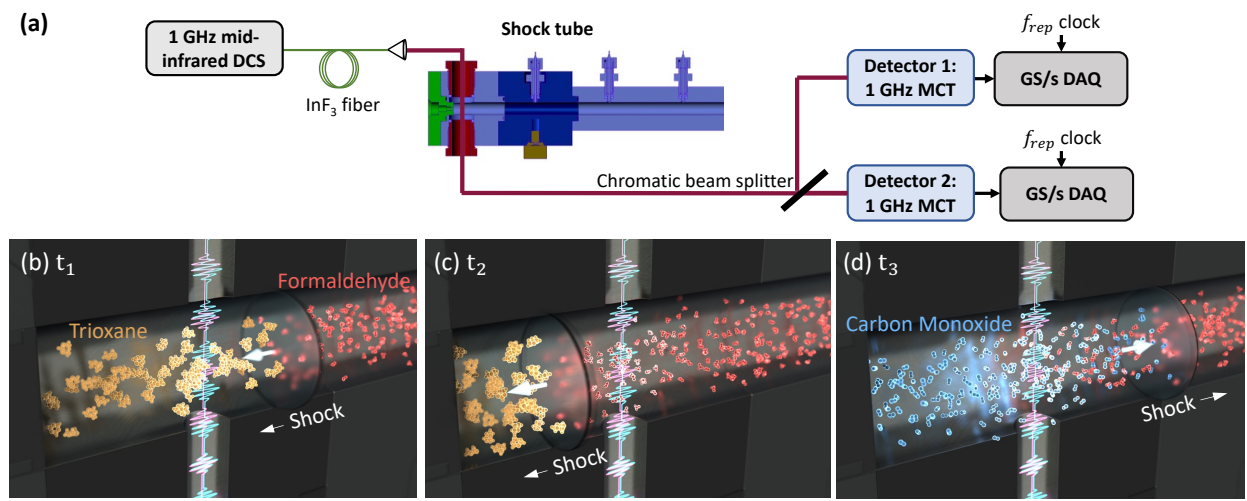

**Figure 2: Experimental Setup**. (a) Experimental schematic. (b-d) Pulses from two mid-infrared frequency combs (colored differently for clarity) are propagated collinearly for dual-comb spectroscopy through optical access close to the shock tube end wall. The dissociation of trioxane (yellow) into intermediate formaldehyde (red) and product carbon monoxide (blue) is initiated by the passage of an incident (b-c) and reflected shock wave (d).

We use a small-diameter high-repetition rate shock tube for the studies (*23*, *32*) (Figure 2). The opening of a high-speed valve separating a low molecular weight high-pressure driver gas and high molecular weight low-pressure driven gas propagates a shock wave down the tube. The shock wave increases the temperature of the driven gas, and after reflecting from the end wall of the tube, propagates through the gas again, further increasing the gas temperature to values as high as 1700 K in these experiments.

We collect an interferogram that represents a full comb-tooth-resolved spectrum every 17.5 μs during 800 consecutive shocks. We ensemble average repeated shock events to optimize signal-to-noise ratio given the short pathlength and small absorbance present in these experiments. Interferograms are aligned and binned relative to the reflected shock, and each bin is phase corrected and averaged together to reach an absorbance noise of ~0.02 in the final averaged spectrum.

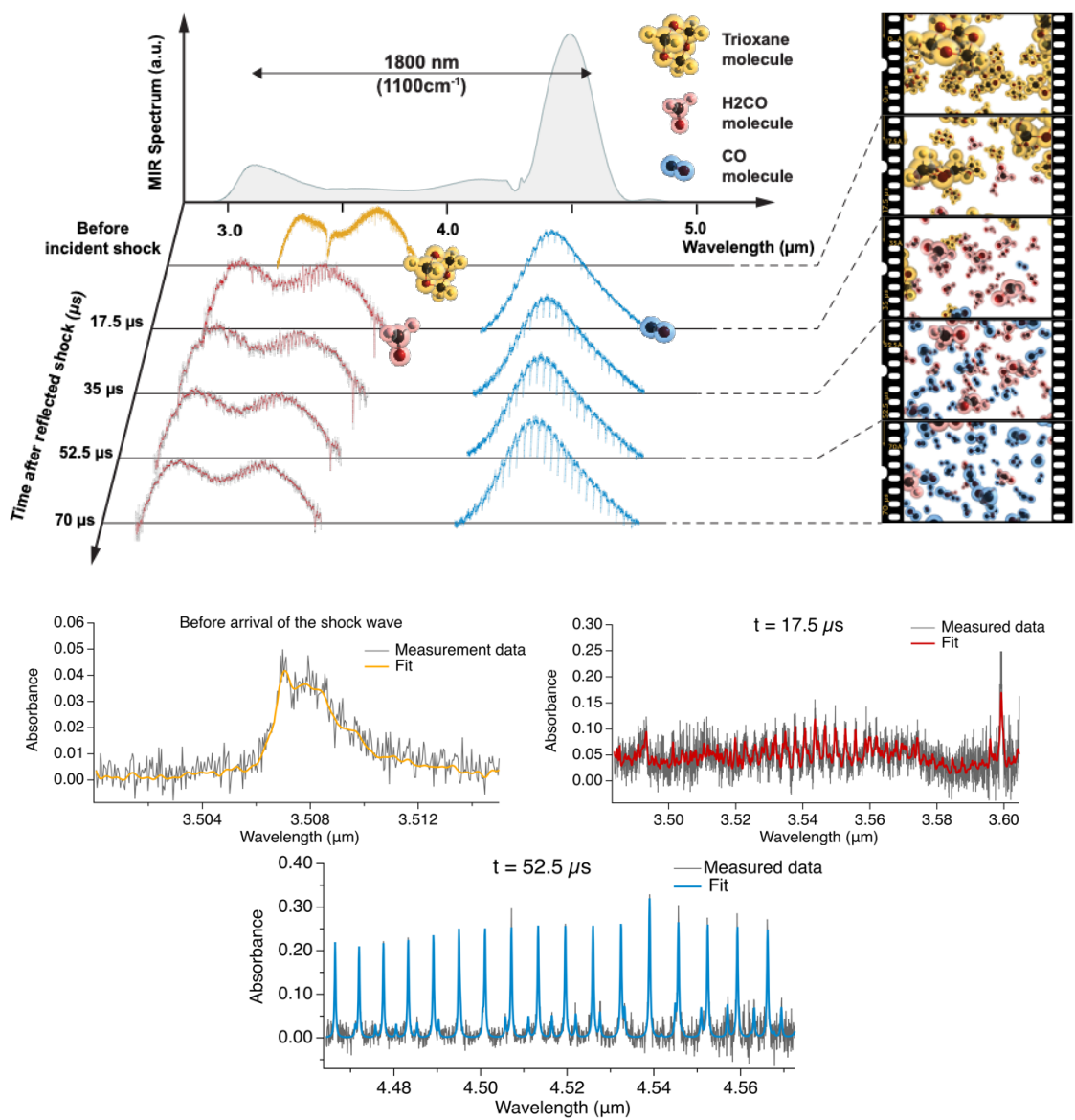

**Figure 3: Breakdown of 1,3,5 Trioxane captured by dual-comb spectroscopy: (top)** Mid-infrared laser output spectrum (linear scale) spanning 1800 nm across the trioxane, formaldehyde and carbon monoxide absorption bands. The two transmission spectra in subsequent frames are recorded with 17.5 μs time steps and are shown after

bandpass filtering and separate detection. The retrieved concentration and temperature create a molecular 'cinema', as the intermediate formaldehyde (detected at $\lambda_c \sim 3.5\ \mu m$) is depleted to form the product carbon monoxide (detected at $\lambda_c \sim 4.5\ \mu m$). **(bottom)** Example of absorption spectra collected for Trioxane before arrival of the reflected shock wave, formaldehyde (17.5 $\mu$s after the reflected shock wave) and carbon monoxide (52.5 $\mu$s after the reflected shock wave) together with fit to the data.

**Results and Discussion**s

We determine gas temperature by comparing the measured CO absorption from 24 rotational transitions of the fundamental vibrational band to the equilibrium Boltzmann distribution. This comparison avoids the influence of lineshape and linestrength errors in molecular absorption databases, which are common at these high temperatures. This method is uniquely available to this spectrometer, which has the optical bandwidth to cover a rotational manifold of CO with high enough resolution to fully resolve each transition.

We quantify species concentrations for CO and $H_2CO$ by fitting a HITRAN2020-based model to the broadband data using the mFID approach (*33*), shown in Figure 3, at the measured rotational temperature (*28*). The HITRAN2020 database (*34*) does not include collision-broadening parameters for CO or $H_2CO$ with argon (which makes up over 90% of the mixture in the shock tube). Additional shock tube measurements were taken of a known mixture of CO (5.02±0.10%) in argon at comparable conditions to the trioxane shocks to assess the impact of the unknown parameters on the retrieved concentration. The spectrometer retrieved a concentration of 5.01±0.33% using HITRAN2020 across all conditions, indicating that there is no discernable concentration bias due to this shortcoming of the database. This resilience to database error also likely stems from the bandwidth of the spectrometer, which allows the fit to optimize over a large number of transitions with different linestrengths rather than just one.

A quantitative reference absorption cross section is not available for trioxane. Instead, 2 ms of DCS data of the stable pre-shock period when trioxane is present and stable were averaged together to create a reference spectrum, and the pre-shock trioxane concentration value was determined from detailed balance between the trioxane concentration before the shock and the CO concentration 1 ms after the shock (when $H_2CO$ was completely consumed). The reference spectrum was then used to fit the time-resolved data during the reaction, creating a relative measurement of trioxane concentration. This limitation can be removed through reference spectroscopy of large hydrocarbons in the future.

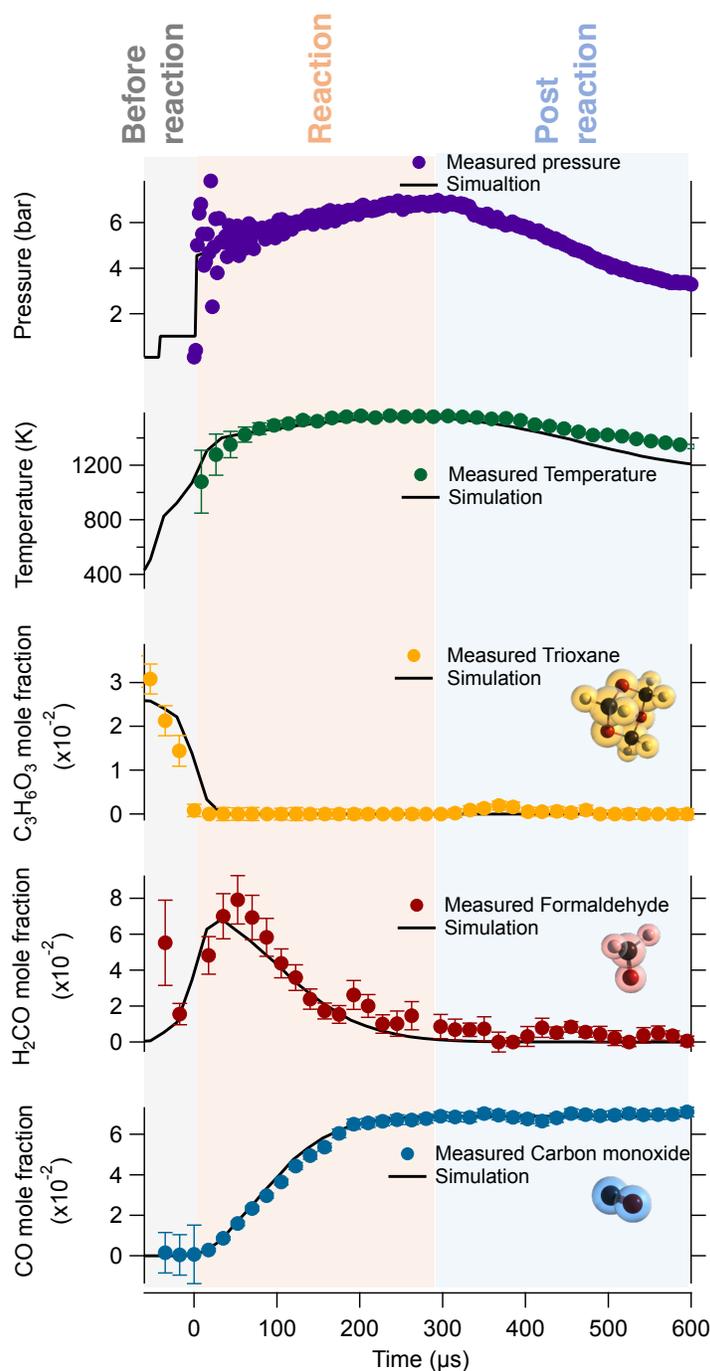

**Figure 4. Reaction time history.** Comparison of DCS measurements and simulations for experiments with 2.6% $C_3H_6O_3$ in argon with reflected shock pressure 4.4 bar. Temperature and mole fractions are 2-point moving averages for clarity. Simulations are based on the kinetic model described in the text, filtered with the same moving average of the data.

We perform four sets of experiments at different temperature, pressure, and precursor conditions to inform the analysis (see supplemental materials). In the experiments, dissociation of trioxane creates formaldehyde, which subsequently dissociates to carbon monoxide. A small amount of iodoethane is added to the initial mixture in some experiments to attempt to isolate H radical abstraction reactions. However, all experiments were already sensitive to bimolecular reactions because of the relatively high initial concentration of trioxane.

Figure 4 shows the measured temperature, concentration, and pressure values from one set of experiments. Typically, in shock tube kinetic studies, reaction rate assignments are made by fitting chemical kinetic models to the composition of a single measured species at a well estimated or measured temperature in extremely dilute conditions. Here, we fit the chemical kinetic model to the concentration of formaldehyde ($H_2CO$), CO, and the spectroscopically measured temperature. This experiment can probe relatively high concentrations of $H_2CO$ because temperature is directly measured. Indeed, the measured temperature ends up significantly lower than the temperature estimated by the frozen normal shock relations. The fit uses a custom pressure-dependent reactor (endwall transducer measured pressure) based on a method employed by Tang and Brezinsky (*35*). The Aramco 3.0 kinetic database (*36*) serves as the base mechanism with the few modifications that are highlighted below and in the supplemental information.

The CO concentration plateaus at t ~ 200 μs, well before the temperature peaks, suggesting complete reaction. The temperature measurement agrees within uncertainty with the modeled temperature trace through the peak temperature until the end of the test time (80% of peak pressure). Excellent agreement between the modeled and experimental $H_2CO$ and CO mole fraction traces is achieved using the updated chemical kinetic model. The chain propagating reaction HCO + M → H + CO + M [(R3) in supplemental information] and chain terminating reaction HCO + H → $H_2$ + CO (R6) of HCO radical are typically very fast and compete, influencing the rate of conversion of $H_2CO$ to CO. The composition of $H_2CO$ and CO is most sensitive to reaction (R3) (Fig 5a.), and its rate is adjusted and reported for all four sets of experiments in Fig. 5b. While high sensitivity does not imply high uncertainty in rate constant assignment, in the case of this system, past measurements of the reaction rate for R3 are uncertain. Our measurements agree with past determinations of the reaction rate for R3 and agree with the rate chosen for the recently reported Aramco 3.0 mechanism (*36*). At our conditions, the self-initiation reaction of $H_2CO$ forming hydroxymethyl ($CH_2OH$) and formyl (HCO) radical is more important in CO formation compared to the study of Friedrichs et al. (*37*) due to the high concentrations of $H_2CO$. Therefore, achieving a better fit to the experimental traces required a faster reaction rate for (R4), closer to that used by Held and Dryer (*38*) in their methanol oxidation mechanism.

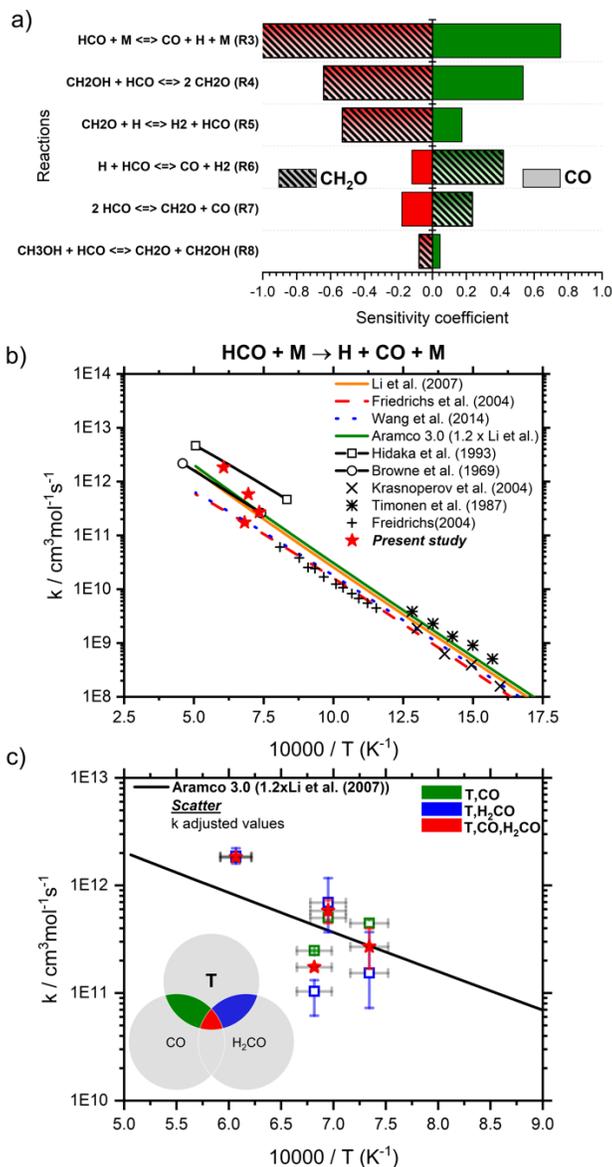

**Figure 5**. a) The sensitivity coefficient of each important reaction in the trioxane decomposition chemical mechanism to formaldehyde ($H_2CO$) and CO composition. The chemical system is most sensitive to the first reaction, and $H_2CO$ and CO composition are strongly correlated (opposite sensitivities), b) The rate constants for $HCO + M \rightarrow H + CO + M$ derived from this work (red stars) compared to the literature, c) The rate constant determined by incorporating different combinations of the measured data into the fits.

Measurements of multiple species can provide additional observables to constrain the series of reactions from precursors to products. Figure 5a shows the reactions in the mechanism that are sensitive to the compositions of the CO and $H_2CO$ observables. The sensitivity analysis shows that these species are negatively correlated in this relatively simple chemical system (i.e. the destruction of $H_2CO$ results in the formation of CO), implying that both molecules carry similar information about the chemical system. Indeed, Figure 5c shows the reaction rate determination based on fitting the model to different combinations of the measured observables. We show that measuring both species and temperature moderately lowers the uncertainty in assigning rate constants for the important reaction R3 (see supplemental).

For many other reactions, where molecular species are less clearly correlated, reaction pathway and rate determinations will benefit more strongly from low uncertainty, quantitative, time-resolved determination of more observables in combination e.g. (*27, 39*).

In addition, there is an ever-growing presence of large data analytic tools such as machine learning that are being applied to complex processes, including chemistry (*40*). Frequency combs offer a bridge between observable-rich mass spectrometry that has been applied in machine learning approaches, and the few species but quantitative and time-resolved nature of laser absorption spectrometry, as recently demonstrated for disease detection (*41*). Achieving both speed and species, this spectrometer opens the door to applying machine learning advances in high-speed chemistry.

**Conclusion**

In summary, we demonstrate a new dual-comb spectrometer that enables in situ speciation and thermometry for microsecond timescale molecular reactions. The spectrometer achieves a balance of high bandwidth, spectral resolution and acquisition rate by leveraging the first 1 GHz repetition rate mid-infrared mode-locked combs. The combs combine chirped pulse amplification with IP-DFG to enable sufficient peak pulse power to generate spectra across more than 1000 cm$^{-1}$ of the chemically important 3-5 μm spectral absorption range. We use the spectrometer to measure the decomposition of trioxane, access conditions sensitive to different formaldehyde chemical reactions compared to other studies, and reliably assign rate constants.

The robust architecture of the frequency combs and motionless design of dual-comb spectrometry represent a powerful platform that should be capable of compact, robust application in field situations in the future, particularly when combined with next generation chip-based frequency combs (*42, 43*). We expect this platform to generate high impact in the fields of physical chemistry, including combustion, hypersonic propulsion and re-entry, planetary science, industrial monitoring, and microscopy.

**Acknowledgement**

**Funding:**
NSF MRI 2019195 (GR, SD, NH)
NSF 2016244 (GR, SD)
NSF 1747774 (PL, MB, RS)
Air Force Office of Scientific Research FA9550-20-1-0328 (GR, NH)
Air Force Research Laboratories FA8650-20-2-2418 (GR, SE, NH)
NIST (SD)

**Author contributions:**
Conceptualization : GR, NH, SD, PL
Data Curation : NH, PC, SE, MB, RS
Formal Analysis : NH, PC, SE, RS, MB
Funding Acquisition: GR, SD, NH
Methodology: NH, PC, GR, PL, RS
Software : PC, SE, NH
Writing: NH, GR, PL, RS, SD

**Competing interests:** Authors declare that they have no competing interests.